\documentclass[12pt]{article}
\usepackage{graphicx}
\usepackage{amssymb}
\newcommand{\Kzs}{\mbox{$\mathrm {K^0_S}$}}

\newcommand{\Jpsi} {\mbox{J\kern-0.05em /\kern-0.05em$\psi$}}

\begin{document}
\title{Heavy Ion physics with the ALICE experiment at the CERN LHC}

\author{J. Schukraft  
\thanks{Talk given at Royal Society meeting on
\it Physics at the high energy frontier --  the Large Hadron Collider project
\rm London, 16 - 17 May 2011, to be published in
\it Philosophical Transactions of the Royal Society A.
} }
\date{CERN, 1211 Geneva, Switzerland}
\maketitle

\begin {abstract}
After close to 20 years of preparation, the dedicated heavy ion experiment ALICE took first data at the CERN LHC accelerator with proton collisions at the end of 2009 and with lead nuclei at the end of 2010. After a short introduction into the physics of ultra-relativistic heavy ion collisions, this article recalls the main design choices made for the detector and summarizes the initial operation and performance of ALICE. Physics results from this first year of operation concentrate on characterizing the global properties of typical, average collisions, both in pp and nucleus-nucleus reactions, in the new energy regime of LHC. The pp results differ, to a varying degree, from most QCD inspired phenomenological models and provide the input needed to fine-tune their parameters. First results from Pb-Pb are broadly consistent with expectations based on lower energy data, indicating that high density matter created at LHC, while much hotter and larger, still behaves like a very strongly interacting, almost perfect liquid. 



\end {abstract}

\section{Introduction}
\subsection{Ultra-relativistic heavy ion physics}

The subject of ultra-relativistic heavy ion physics is the study of strongly interacting matter under extreme conditions of high temperature and/or high matter density. The {\it matter} under study is the one which fills the Universe today with the elements of the periodic table, essentially made from quarks, the basic building blocks of matter, and held together by gluons, the carriers of the strong (nuclear) force. Under normal conditions, i.e. the ones currently prevailing in the Universe, the quarks are bound ('confined') into composite objects called baryons (bound states of three quarks, with the proton and neutron being the most prominent examples) and mesons (bound states of a quark and anti-quark pair). Quantum Chromodynamics (QCD), the theory of strong interactions, predicts that at a sufficiently high energy density there will be a transition from ordinary nuclear or hadronic matter to a plasma of free ('deconfined') quarks and gluons -- a transition which took place in the early Universe a few microseconds after the Big Bang and which might still play a role today in compact stellar objects. The discovery and characterisation of this new high temperature phase of matter, called the 'Quark-Gluon Plasma' (QGP), requires a sufficiently large volume of hot matter and is therefore pursued in collisions of heavy nuclei at the highest possible energy.

The Quark-Gluon Plasma is thought to be the ground state of QCD, where quarks and gluons are deconfined and chiral symmetry is (approximately) restored, i.e. the quarks are no longer bound into composite particles (collectively called 'hadrons') and are approximately massless, retaining only the bare mass associated with the Higgs mechanism. 
The transition from normal matter to the QGP is expected to happen at a critical temperature of the order of the QCD energy scale parameter $\Lambda_{QCD}$, which is $\approx$ 200 MeV (more than $10^{12}$ Kelvin).
At such an energy scale, the standard method of solving the equations of QCD by pertubative approximations is not very reliable. The theoretical description is therefore in need of -- as well as a unique testing ground for -- new approaches to QCD in the regime where the strong interaction is truly strong. A large number of these have been brought to bear on the subject, from numerical solutions on large computers ('lattice QCD'), statistical and thermodynamical formulations, classical solutions in the high density limit ('Colour Glass Condensate'), up to string theory and quantum gravity ('Conformal Field Theory in Anti-de Sitter Space or AdS/CFT').

This study of the phases of nuclear matter is relevant also beyond the specific context of QCD, as  phase transitions and symmetry breaking are central concepts in the Standard Model of particle physics and the predicted QCD transition is the only one directly accessible to laboratory experiments. It may also provide some insight into the properties of large and complex systems involving elementary quantum fields, indicating how the microscopic laws of physics (the 'QCD equations') give rise to macroscopic phenomena such as phase transitions and critical behaviour.

Using methods and
concepts from both nuclear and high-energy physics, it constitutes a
new and interdisciplinary approach in investigating matter and its
interactions. In high-energy physics, {\it interactions} are derived from 
first principles (gauge theories), and the {\it matter} concerned consists mostly
of single particles (composite or elementary). In contrast, on nuclear physics scales the strong {\it interaction} is shielded and can, to date, only be
described in terms of effective theories, whereas the {\it matter} under investigation consists of extended
and interacting systems with collective many-body features. 
Ultra-relativistic heavy ion physics combines the {\it elementary-interaction} aspect of
high-energy physics with the {\it macroscopic-matter} aspect of nuclear physics.

\subsection{Heavy ion physics at lower energies}
The study of heavy-ion collisions is a rapidly evolving field. After the pioneering experiments at the LBNL in Berkeley and the JINR in Dubna in the 70's with relativistic heavy ions
(Energy/rest mass $E/m \approx 1)$, the first experiments at ultra-relativistic energies $(E/m \gg 1)$ started in 1986 with light ions
almost simultaneously at the Brookhaven AGS and the CERN SPS fixed target accelerators. Really heavy ions (A $\approx$ 200) have been available in the AGS since the end of 1992 and at the SPS since the end of 1994.

The turn of the century (2000) was an extremely important milestone. It started with an appraisal of the CERN SPS Pb beam results \cite{Heinz:2000bk} which concluded that 'compelling evidence has been found for a new state of matter' featuring many of the characteristics expected for a Quark-Gluon Plasma. Later that year the Relativistic Heavy Ion Collider (RHIC) and its four experiments at BNL began operation with Au-Au collisions at 130 GeV/nucleon. Already the first short run produced a wealth of interesting and novel data, and RHIC has kept producing exciting and at times surprising results ever since. A 'Heavy Ion Standard Model' (HISM) began to emerge which characterises the new high density state created in high energy nuclear collisions as an extremely strongly interacting and almost perfect fluid \cite{rhichwhitepaper}, sometimes called the 'sQGP' (where the 's' stands for 'strongly interacting'). It is almost opaque and absorbs much of the energy of any fast parton (quark or gluon) which travels through this matter -- a process referred to as 'jet quenching\footnote{Because of confinement, even a fast (high momentum) parton cannot be observed directly. It will fragment into a number of hadrons which can be detected as a narrow 'jet' of particles around the original parton direction.}' -- and it reacts to pressure gradients by flowing almost unimpeded and with very little internal friction (i.e. very small viscosity).  

When the LHC came into operation with ion beams at the end of last year (November 2010), the available energy in the centre of mass system has increased by four orders of magnitude in a little over 25 years \cite{JSBromley}.
This rapid progress was only possible by reusing accelerators, and initially even detectors, built over a longer time scale for Particle Physics; RHIC remains the only dedicated facility.  Today, with more than 2000 physicists active worldwide in this field, ultra-relativistic heavy ion physics has moved, in less than a generation, from the periphery into a central activity of contemporary Nuclear Physics.

\section{The ALICE experiment}

\subsection{Designing ALICE}

ALICE, which stands for A Large Ion Collider Experiment, is very different in both design and purpose from the other experiments at the LHC~\cite{Schukraft:2010rt,Evans:2009zz}. It is specifically optimized to study heavy ion reactions, but data taking with pp (and later p-nucleus) is equally part of the programme primarily to collect comparison data for heavy ions~\cite{Carminati:2004fp,Alessandro:2006yt}.

ALICE was 'born' in December 1990, when a first meeting took place between a few dozen physicists to contemplate how, and in fact if, to build a detector dedicated to heavy ion physics at the LHC.
Designing a dedicated heavy ion experiment in the early 90's for use almost 20 years later posed some significant challenges: In a field still in its infancy -- with the SPS lead programme yet to start -- it required extrapolating the conditions to be expected by a factor of 300 in energy and a factor of 7 in beam mass. The detector therefore had to be both 'general purpose' -- able to measure most signals of potential interest, even if their relevance may only become apparent later -- and also flexible, allowing additions and modifications along the way as new avenues of investigation would open up. In both respects ALICE did very well, as it included a number of observables in its initial physics menu whose importance only became clear after results appeared from RHIC (e.g. reconstructing secondary decay vertices for heavy quarks, particle identification up to large transverse momentum), and various major detection systems were added to the experiment over time to match the evolving physics goals, from the muon spectrometer in 1995, the transition radiation detector (TRD) in 1999, to a large jet calorimeter (EMCAL) added as recently as 2008.

Other challenges relate to the experimental conditions expected for nucleus-nucleus collisions at the LHC. The most difficult one to meet is the extreme number of particles produced in central head-on collisions, which at the time were thought to be up to three orders of magnitude larger than in typical proton-proton interactions at the same energy, and a factor two to five above the highest multiplicities expected at RHIC. The tracking of these particles was therefore made particularly safe and robust by using mostly three-dimensional hit information with many points (up to 200) along each track  in a moderate magnetic field (B = 0.5 Tesla) to ease the problem of recognizing and measuring charged particles.
In addition, a large dynamic range is required for momentum measurement, spanning more than three orders of magnitude from tens of MeV to well over 100 GeV. This is achieved with a combination of detectors with very low material thickness (to reduce scattering of low momentum particles) and tracking particles over a large distance $L$ of up to 3.5 m, which gives a figure of merit for momentum resolution, $BL^2$, quite comparable to those of the other LHC experiments. In addition, the vertex detector with its six silicon planes, four with analogue read-out, can be used as a standalone spectrometer with momentum and particle type information to measure particles that do not reach the outer tracking detectors.

Finally, Particle Identification (PID) over much of this momentum range is essential, as many phenomena depend critically on either particle mass or particle type. ALICE therefore employs essentially all known PID techniques in a single experiment, including energy loss in silicon and gas detectors, Cherenkov and transition radiation, time-of-flight, electromagnetic calorimeters, as well as topological decay reconstruction.

As the intensity of heavy ion beams in the LHC is rather modest, with interaction rates of order 10 kHz or less with Pb beams, rather slow detectors can be employed like the Time Projection Chamber (TPC) and silicon drift detectors. Only moderate radiation hard electronics and trigger selectivity are required and most of the read-out is sequential, not pipelined (i.e. only one single event can be processed at a time in ALICE). However, because the amount of information produced in heavy ion interactions is huge (up to 100 Mbyte/event) and the statistics has to be collected in a short time (the LHC runs with ions only one month every year), the Data Acquisition System has been designed for a very high bandwidth of over 1 Gbyte/s to permanent storage, larger than the throughput of all other LHC experiments combined. 

The ALICE design evolved from the Expression of Interest (1992) via a Letter of Intent (1993) to the Technical Proposal (1996) and was officially approved in 1997.  The first ten years were spent on design and an extensive R\&D effort. It became clear from the outset that the challenges of heavy ion physics at the LHC could neither be met nor paid for with existing technology. Significant advances, and in some cases a technological break-through, would be required to actually build what physicists had dreamed up on paper for their experiments. The initially very broad and later more focused, well organised and well supported R\&D effort, which was sustained over most of the 1990's, has lead to many evolutionary and some revolutionary advances in detectors, electronics and computing~\cite{Evans:2009zz}. 

\subsection{The ALICE detector}

\begin{figure}[!t]

\centerline{\includegraphics[width=1.0\textwidth]{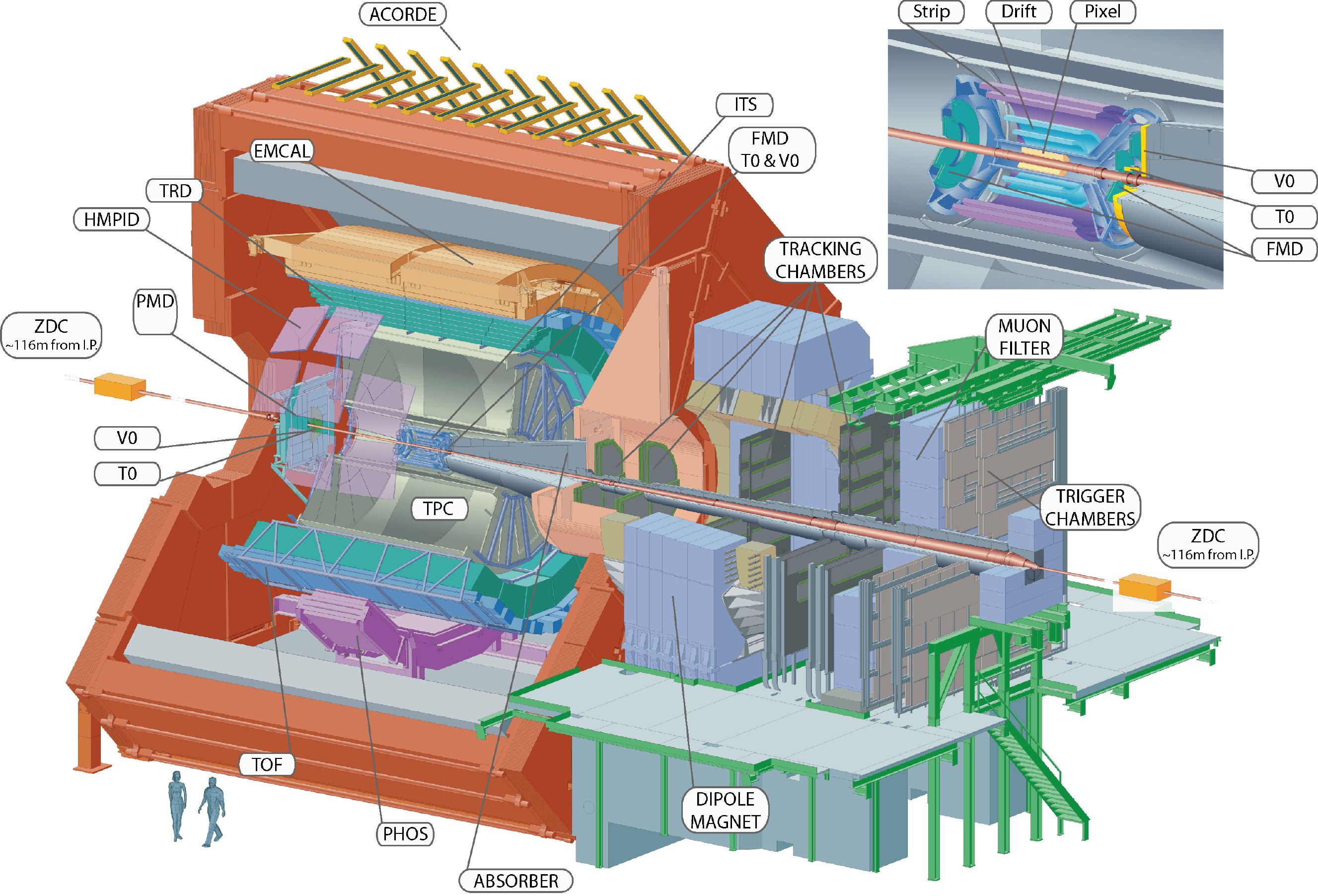}}
\caption{ALICE uses 18 different detector systems, indicated in the figure with their acronyms. The central detectors are located inside a large solenoid magnet (left side), the forward Muon arm spectrometer with its dipole magnet is located on the right. The insert top right is a blow-up of the interaction region showing the 6 layers of the silicon vertex detector (ITS) and the forward trigger and multiplicity detectors (T0, V0, FMD).
}
\label{setup}
\end{figure}

ALICE is usually referred to as one of the smaller detectors, but the meaning of 'small' is very relative in the context of LHC:  The detector stands 16 metres tall, is 16 m wide and 26 m long, and weights approximately 10,000 tons. It was designed and built over almost two decades by a collaboration which currently includes over 1000 scientists and engineers from more than 100 institutes in some 30 different countries. The experiment consists of 18 different detection systems, each with its own specific technology choice and design constraints. 
A schematic view of ALICE is shown in Figure~\ref{setup}. It consists of a central part, which measures hadrons, electrons, and photons, and a forward single arm spectrometer that focuses on muon detection. The central 'barrel' part covers the direction perpendicular to the beam from $45^0$ to $135^0$ and is located inside a huge solenoid magnet, which was built in the 1980's for the L3 experiment at CERN's LEP accelerator. As a warm resistive magnet, the maximum field at the nominal power of 4 MW reaches 0.5 T. The central barrel contains a set of tracking detectors, which record the momentum of the charged particles by measuring their curved path inside the magnetic field. These particles are then identified according to mass and particle type by a set of particle identification detectors, followed by two types of electromagnetic calorimeters for photon and jet measurements. The forward muon arm ($2^0-9^0$) consists of a complex arrangement of absorbers, a large dipole magnet, and fourteen planes of tracking and triggering chambers.

{\it Tracking detectors:} 
Tracking in the central barrel is divided into the Inner Tracking System (ITS), a six-layer, silicon vertex detector, and the Time-Projection Chamber (TPC). The ITS locates the primary vertex (where the interaction took place) and secondary vertices (decay points of unstable particles) with a precision of order a few tens of micrometers. Because of the high particle density, the innermost four layers need to be high resolution devices, i.e. silicon pixel and silicon drift detectors, which record both x and y coordinates for each particle hit (the 3rd coordinate being fixed by their radial position in the experiment). The outer layers are equipped with double-sided silicon micro-strip detectors. The total area covered with silicon detectors reaches 7m$^2$ and includes almost 13 million individual electronic channels.

The need for efficient and robust tracking has led to the choice of a TPC as the main tracking detector. In spite of its drawbacks concerning slow recording speed and huge data volume, a TPC can provide reliable and redundant momentum and PID information. The large TPC (5 m length, 5.6 m diameter) has been specifically optimized for heavy ion physics in terms of very low mass and good performance up to extremely high particle densities.

{\it Particle identification detectors:}
Particle identification with large acceptance and for many different particles is an important design feature of ALICE with several detector systems dedicated to PID: the Time-of-Flight (TOF) array measures the flight time of particles from the collision point out to the detector; together with the momentum this time determines the mass. It covers the central barrel over an area of 140 m$^2$ with 150,000 individual cells at a radius of close to 4 m. The requirement for an affordable system with a large number of channels, as well as state-of-the-art time resolution of better than 100 ps, was solved with the development of a novel type of gas detector, the Multigap Resistive Plate Chamber. 
The TOF method, which is limited to lower energy particles with a Lorentz factor $E/m \lesssim 3-6$, is complemented by two technologies which measure the electromagnetic radiation emitted by highly relativistic particles which travel through media: Cherenkov radiation in the eV energy range and transition radiation in the keV range.
The HMPID detector is a single-arm, 10 m$^2$ array of ring imaging Cherenkov counters with liquid radiator and solid CsI photocathode evaporated on the segmented cathode of multiwire proportional chambers. It extends the particle identification capabilities toward higher momenta in about 10\% of the barrel acceptance. The Transition Radiation Detector (TRD) will identify electrons above 1 GeV to study production rates of heavy quarks (charm and beauty mesons). It consists of six layers of Xe/CO$_2$ filled time expansion chambers following a composite foam and fibre radiator and includes a distributed tracking processor in the front-end electronics to select (i.e. trigger on) interesting events in real time. 

{\it Calorimeters:}
Photons, spanning the range from thermal emission to hard QCD processes, as well as neutral mesons are measured in the small acceptance, high-resolution and high-granularity PHOS electromagnetic calorimeter. It is located far from the vertex (4.6 m) and made of dense scintillating crystals (PbWO$_4$) in order to cope with the large particle density. A set of multiwire chambers (CPV) in front of PHOS helps with discriminating between charged and neutral particles.
The interaction and energy loss of high-energy partons in dense matter will play an essential role in the study of nuclear collisions at the LHC. In order to enhance the capabilities for measuring jet properties, a second electromagnetic calorimeter (EMCal) was installed in ALICE by end 2010.  The EMCal is a Pb-scintillator sampling calorimeter with longitudinal wavelength-shifting fibres, read out via avalanche photo diodes. Much larger than PHOS, but with lower granularity and energy resolution, it is optimized to measure jet production rates and jet characteristics in conjunction with the charged particle tracking in the other barrel detectors. 

{\it Forward and trigger detectors:}
A number of small and specialized detector systems are used for event selection or to measure global features of the reactions. 
The collision time is measured with extreme precision ($<$ 20 ps) by the 'T0' detector; two sets of 12 Cherenkov counters (fine mesh photomultipliers with fused quartz radiator) mounted tightly around the beam pipe. Two arrays of segmented scintillator counters, called 'V0' detector, are used to select interactions and to reject beam related background events. An array of 60 large scintillators (ACORDE) on top of the L3 magnet triggers on cosmic rays for calibration and alignment purposes, as well as for cosmic ray physics.
The Forward Multiplicity Detector (FMD) provides information about the number and distribution of charged particles emerging from the reaction over an extended region, down to very small angles.  These particles are counted in rings of silicon strip detectors located at three different positions along the beam pipe.
The Photon Multiplicity Detector (PMD) measures the multiplicity and spatial distribution of photons in each single heavy ion collision. It consists of two planes of gas proportional counters with cellular honeycomb structure, preceded by two lead plates to convert the photons into electron pairs. 
Two sets of small, very compact calorimeters (ZDC) are located far inside the LHC machine tunnel ($>$ 100 m) and very close to the beam direction to record neutral particles which emerge from heavy ion collisions in the forward direction.

\begin{figure}[!t]

\centerline{\includegraphics[width=1.0\textwidth]{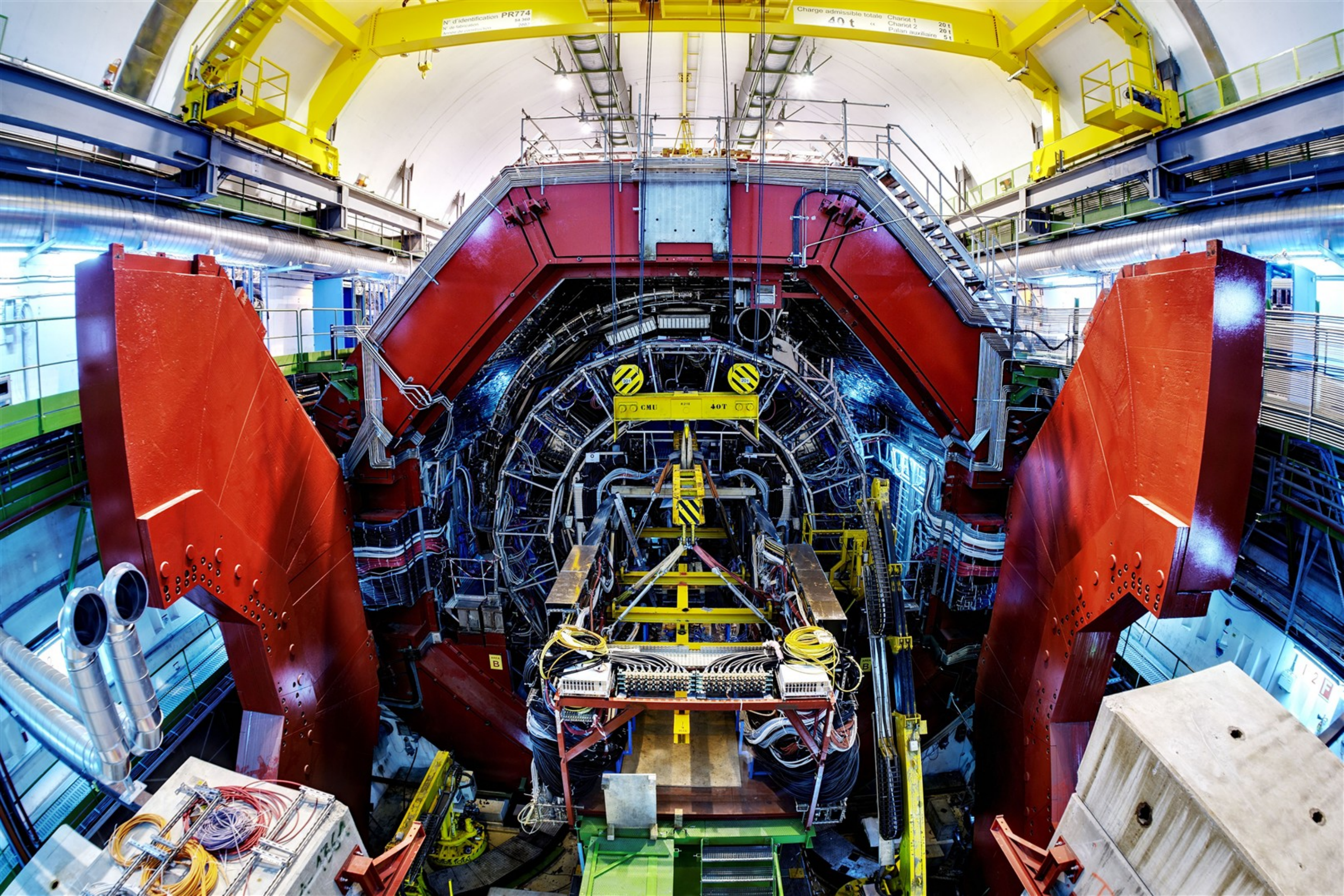}}
\caption{Front view of the ALICE magnet, with its doors fully open. A bridge guides services, power cables, fluids and gases into the central detector. The silicon tracker is no longer visible and even the large TPC is mostly obscured. The stainless steel space-frame structure which supports most of the central detectors is partially filled with EMCAL, TOF, and TRD modules.}
\label{alicephoto}
\end{figure}

Figure~\ref{alicephoto} shows the front of the ALICE barrel detectors and the magnet during detector installation in 2009.

{\it Muon Spectrometer:}
 On one side behind the central solenoid and at small angles between $2^0$ and $9^0$ relative to the beam direction is a muon spectrometer which includes a dipole magnet generating a maximum field of 0.7 T. Several passive absorber systems (a hadron absorber close to the interaction point, a lead-steel-tungsten shield around the beam pipe and an iron wall) shield the spectrometer from most of the reaction products. The penetrating muons are measured in 10 planes of cathode pad tracking chambers, located between 5m and 14 m from the interaction, with a precision of better than $ 100 \mu$m. Again the relative low momentum of the muons of interest and the high particle density are the main challenge; therefore each chamber has two cathode planes, which are both read out to provide two-dimensional space information, and the chambers are made extremely thin and without metallic frames. The individual cathode pads range in size from 25 mm$^2$ close to the beam up to 5 cm$^2$ further away and cover 100 m$^2$ of active area with over 1 million active channels. Four trigger chambers are located at the end of the spectrometer, behind a 300 ton iron wall to select and trigger on pairs of muons, including those from the decay of heavy quark particles (of main interest for ALICE are the $\Jpsi$ and $\Upsilon$ mesons). The chambers are made in the 'resistive plate' technology widely used by LHC experiments, and of modest granularity (20,000 channels covering 140 m$^2$).  
 
The layout of the ALICE detector and its various subsystems is described in more detail in~\cite{Evans:2009zz, ALICEdet}.

\section{Initial operation and performance}
The very first pp collisions were observed in ALICE on $\rm 23^{rd}$ November  2009, when the LHC, during the very early commissioning phase, provided an hour of colliding beams for each of its four large experiments~\cite{Schukraft:2010rt}. Such was the enthusiasm about 'real data', after years of simulation exercises, that this first harvest of some 300 events, significantly less than the number of ALICE collaborators, was analysed right away and made it into a physics publication only five days later~\cite{:2009dt}; well before stable beams were declared on December 6 and sustained data taking could start.  

The many years of preparation, analysis tuning with simulations, and detector commissioning with cosmic rays during much of 2008/9 paid off quickly with most of the detector components working 'out of the box' and rather close to performance specifications. Within a few days all experiments were showing first qualitative results and the first phase of LHC physics, often referred to as the 'rediscovery of the standard model', was getting under way~\cite{LHCstatrep}. The various members of the 'particle zoo' created in pp collisions made their appearance in ALICE in rapid succession, from the easy ones 
($\pi, K, p, \Lambda, \Xi, \phi,..$) in 2009 to the more elusive ones  when larger data sets were accumulated early 2010 (K*,$ \Omega$, charmed mesons, $J/\Psi$, ..). 

However, precise results and small systematic errors need more than large statistics and a good detector performance; they require a precise knowledge and understanding of the detector as well. The next months were therefore spent on 'getting to know' the experiment in greater detail, including calibration, alignment, material distribution and detector response which are all crucial ingredients for the analysis and correction procedures. 
For example, the amount of material of the detector 'as built' (rather than 'as designed') had to be measured with the help of photons that convert in the material into electron-positron pairs. With this method, the material thickness was determined to about 5\% relative accuracy. Such a precision is important e.g. for the measurement of the antiproton to proton ratio, because annihilation of antiprotons in the detector material is one of the dominating factors of the systematic error.

\begin{figure}[!t]
\centerline{\includegraphics[width=1.0\textwidth]{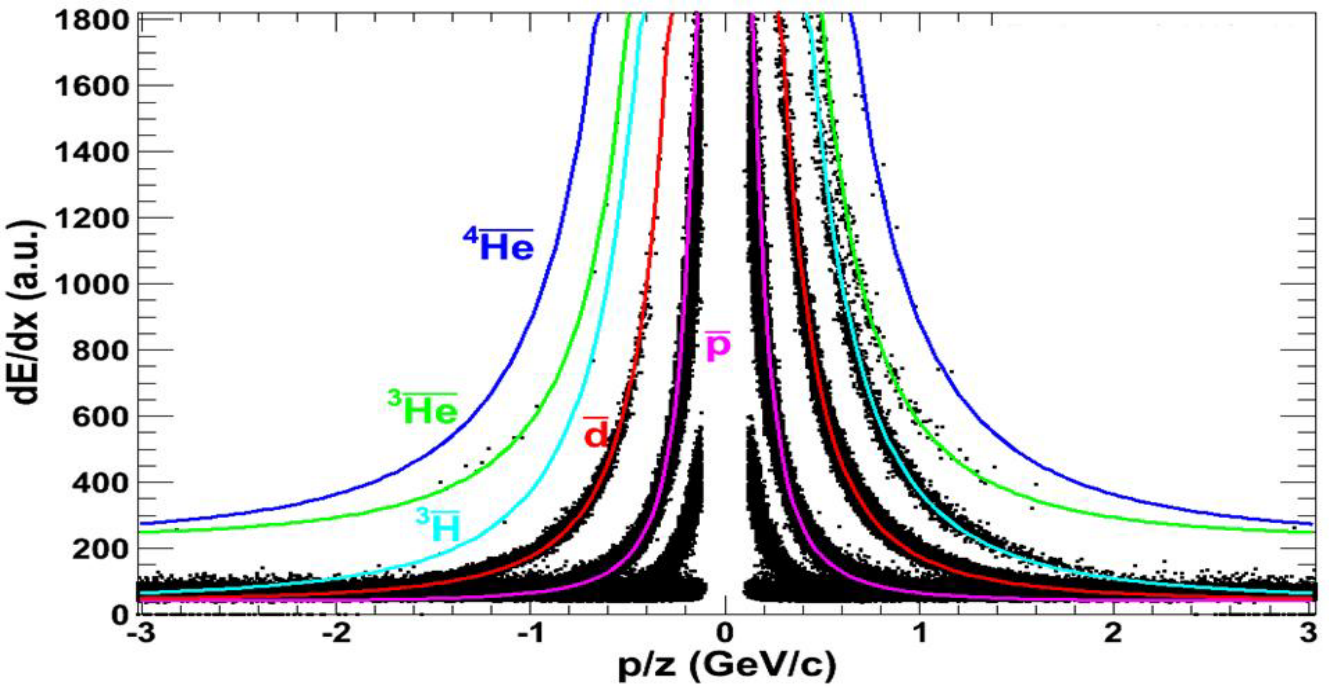}}
\caption{ Energy loss distribution versus rigidity (momentum/charge) for primary and secondary particles reaching the TPC. The lines overlaid on the distribution correspond to the expected energy loss for different particles and nuclei.}
\label{TPCdedx}
\end{figure}

Measuring the precise position of the many millions of detection elements inside the experiment is a crucial pre-requisite for accurate momentum determination. To improve on the initial information provided by geometrical survey, this 'detector alignment' started with cosmic rays and continues to date with beam, quickly reaching an accuracy of $< 10 \mu$m for the innermost layers of the vertex detector, about 200-300 $\mu$m for the TPC, and the millimetre level in the outer detectors required for track matching.
Likewise, accurate gain and pulse height calibrations are needed in particular for the particle identification using energy loss (dE/dx). The energy loss distribution in the TPC, which has reached its design resolution of better than $6\%$ relative accuracy in dE/dx for long tracks, is shown in Figure~\ref{TPCdedx} versus rigidity, separately for positive and negative charges. Different particle species and light nuclei are clearly separated in the non-relativistic momentum region. Note that the apparent asymmetry between nuclei and anti-nuclei production rates in this plot is an instrumental effect. Here tracks are not required to point precisely back towards the vertex and therefore many secondary particles knocked out from the detector material are included. Even in proton-proton collisions, anti-nuclei as heavy as anti-tritium can be produced. In heavy ion reactions, even a handful of anti-Helium candidates, the heaviest anti-nuclei ever produced in the laboratory, have been identified by a combined energy loss and time of flight measurement.  

\section{Results with proton-proton collisions}

Data taking with proton beams was focused during most of 2010 on collecting a large sample ($> 10^9$) of average (so called 'minimum bias') collisions and to measure their global event properties, which are needed for comparison to the heavy ion data~\cite{Schukraft:2010ru}.
Both the average number of charged particles (${\rm d}N_{\rm ch}/{\rm d}\eta$) as well as their multiplicity distributions in different acceptance windows were measured at 0.9, 2.36 and 7 TeV for  inelastic and non-single diffractive collisions~\cite{:2009dt,Aamodt:2010pp,Aamodt:2010ft}. 
The dependence of the average multiplicity on the centre-of-mass energy $\sqrt{s}$ is well described by a power law, $s^{0.11}$, and increases significantly faster than predicted by most QCD based simulation programs ('event generators'). Most of this stronger increase happens in the tail of the multiplicity distribution, i.e. for events with much larger than average multiplicity. Likewise, neither the transverse momentum ($p_T$) distribution at 900 GeV nor the dependence of average $p_T$ on charged particle multiplicity is well described by these generators~\cite{Aamodt:2010my}, in particular when including low momentum particles ($p_T < 500$ MeV). The shape of the $p_T$ spectrum as a function of multiplicity hardly changes below  0.8 GeV (which includes the large majority of all particles), whereas the high momentum power law tail typical of hard QCD scattering increases rapidly with event multiplicity above about 1-2 GeV.

The yields and $p_T$ spectra of identified stable charged ($\pi$, K, p) and neutral strange particles 
($\Kzs,\Lambda, \Xi, \phi$) have been measured for the small 900 GeV data set taken in 2009~\cite{Aamodt:2011zj,Collaboration:2010vf}. Most event generators predict yields well below the data -- by factors of two to almost five -- and more so at high $p_T$ and for the heavier particles ($\Lambda, \Xi$). The ratio of $\Lambda$ to $\Kzs$ agrees very well with pp data at 200 GeV from RHIC but is significantly below the ratio measured at the p-pbar colliders at CERN and Fermilab. This discrepancy merits further investigation; it could be due to differences between the experiments in the acceptance, triggers, or correction for feed-down from weak interaction decays. The relative production ratios of baryons to mesons (e.g. the proton-to-pion ratio) are of particular interest, as they rise well above unity in nuclear collisions at RHIC. This 'meson-baryon' anomaly has been interpreted as an indirect sign of the QGP, in which case it would not be obvious why similar ratios should be reached already in minimum bias pp interactions.

Baryon number (which counts the number of baryons minus anti-baryons) is a conserved quantity in the high energy physics Standard Model because quarks, which make up baryons, are created (and annihilated) always in particle anti-particle pairs.  
In high energy proton-proton collisions (two incoming baryons), the projectiles in general break up into several hadrons including two excess baryons (which may or may not be protons). The outgoing baryons retain only a fraction of the original beam energy and emerge under a finite angle (finite rapidity\footnote{Rapidity ($y$) is a relativistic and convenient measure of the polar angle, invariant under Lorentz boost along the beam direction. For massless (or highly relativistic) particles it coincides with the pseudorapidity($\eta$), defined as $\eta =$  -ln  tan $ \theta/2$, where $\theta$ is the polar angle with respect to the beam line.}
difference) with respect to the original beam particles. The deflection and deceleration (or 'baryon stopping') of the incoming projectile, or more precisely of the conserved baryon number associated with the beam particles, is often called 'baryon-number transport'.
At the LHC, by far the highest energy proton-proton collider, ALICE has studied the distribution of net baryon number along the beam direction over very large rapidity intervals (large deflection/large deceleration)
by measuring the antiproton-to-proton ratio at mid-rapidity~\cite{Aamodt:2010dx} (i.e. perpendicular to the beam) in order to discriminate between various theoretical models of baryon stopping. Baryon number transport over large distances in rapidity is often described in terms of a nonlinear three gluon configuration called 'baryon string junction'. The dependence of this process on the size of the rapidity gap has been a longstanding issue (for large gaps, where it should be dominant), with advocates for both very weak and rather strong dependencies.
In either case, the $\bar{p}/p$ ratio at LHC is expected to be very close to unity, with the difference between various models only of the order of a few percent. So this ratio must be measured with high precision. While in the ratio many instrumental effects cancel, the very large difference between $p$ and $\bar{p}$ cross-section for both elastic (track changes direction and can get lost) and inelastic (particle can be absorbed)  reactions with the detector material lead to corrections of order 10\%, even in the very thin central part of ALICE. As the corrections are much larger than the effect, a very precise knowledge of the detector material as well as of the relevant cross-section values at low momentum is required. The former was measured with the data via photon conversions; the latter had to be cross-checked with experimental data since all available versions of the standard computer program used to simulate detector response overestimated the $\bar{p}$ cross-sections by up to a factor of five. The $\bar{p}/p$ ratio was found to be compatible with 1.0 at 7 TeV and 4\% below 1.0 at 900 GeV, with an experimental uncertainty of about 1.5\%, dominated by the systematic error. This result favours models which predict a strong suppression of baryon transport over large gaps; they agree very well with standard event generators but not with those that have implemented enhanced proton stopping.

\section{Results with Lead-Lead collisions}

\begin{figure}[!t]
\centerline{\includegraphics[width=1.0\textwidth]{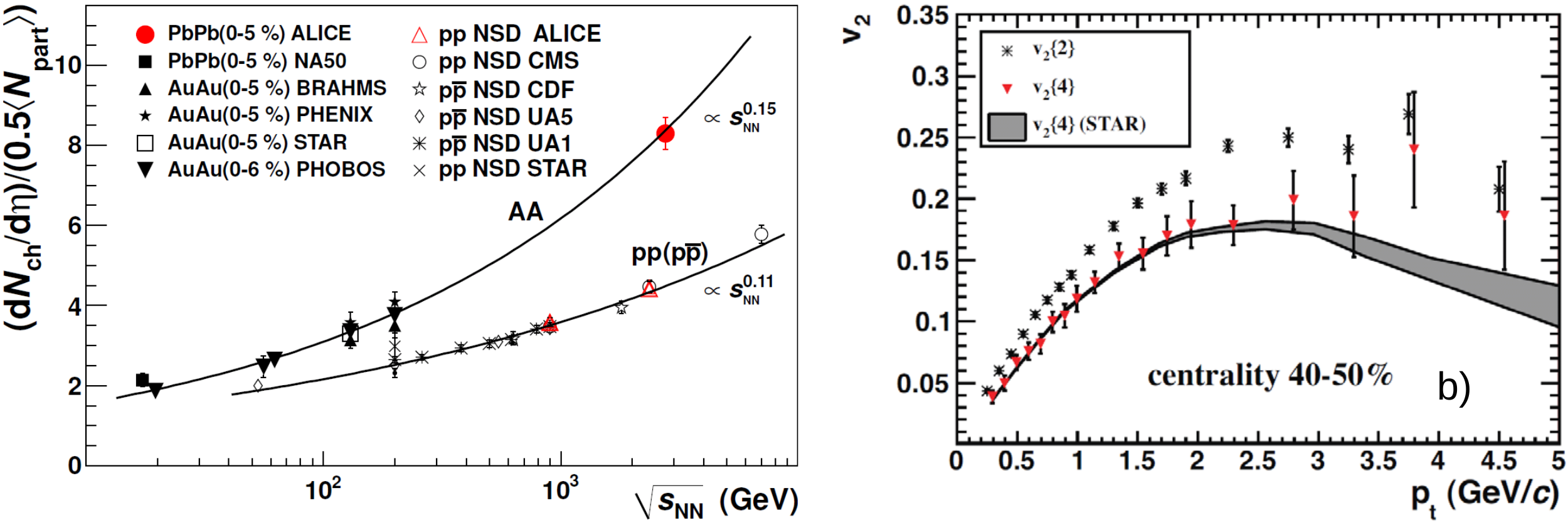}}

\caption{Left: Charged particle pseudorapidity density ${\rm d}N_{\rm ch}/{\rm d}\eta$ per colliding nucleon pair ($0.5 N_{\rm part}$) versus centre of mass energy for pp and AA collisions. Right: Elliptic flow coefficient $v_2$ versus transverse momentum $p_T$ compared to RHIC data from the STAR experiment. Figures taken from refs~\cite{Aamodt:2010pb, Aamodt:2010pa}. }
\label{figpb1}
\end{figure}

With the fantastic results from RHIC and the emergence of the Heavy Ion Standard Model (HISM) mentioned earlier, the {\em search} for the QGP is essentially over and the {\em characterisation} of the new high temperature phase is well under way. However, {\em precision measurements} of QGP parameters are just starting (precision is meant here in the context of non-pertubative QCD, where a factor of two is already respectable). At up to 30 times higher energy, the LHC has some unique advantages compared to RHIC and is complementary in other aspects. Significant differences  are expected at the higher energy in terms of energy density, lifetime and volume of the state of matter created in the collisions, and rare 'hard probes' (jets, heavy flavours) will be produced abundantly. In this new environment, the HISM can be tested and validated. Once it has been verified that the global event characteristics (e.g. energy density, volume, lifetime) of matter at LHC are indeed rather different, but that evolution and intrinsic properties are still well described  by the HISM, one can embark on the programme of precision measurements of the QGP parameters (e.g. viscosity, equation-of-state, transport coefficients, Debye screening mass,..)~\cite{Schukraft:2011kc}. 

The first, and most anticipated, results concerned global event characteristics. When ALICE was conceived in the early '90s, the charged particle density predicted for 
central Pb-Pb collisions at LHC was extremely uncertain, varying between 2000 and $>4000$ charge particles per unit rapidity. With RHIC results, the uncertainties came down substantially, as did the central value, with most predictions clustering in the range 1000 - 1700~\cite{Abreu:2007kv}.
The value finally measured at LHC, ${\rm d}N_{\rm ch}/{\rm d}\eta \approx 1600$~\cite{Aamodt:2010pb}, was right in this range, if somewhat on the high side. 
From the measured multiplicity in central collisions one can derive a rough estimate of the energy density, which gives at least a factor three above RHIC; the corresponding increase of the initial temperature is about 30\%.
This is indeed matter under extreme conditions, reaching a temperature which is 200,000 times hotter than the centre of our sun and an energy density fifty times larger than in the core of a neutron star!

When combined with lower energy data, the charged particle production per participant\footnote{Various 
notations and measures are used to classify nuclear collisions: In 'central' collisions, the two nuclei collide head-on (impact parameter close to zero), in 'peripheral' ones they only touch at the edges (impact parameter close to twice the nuclear radius). The collision geometry is captured more quantitatively by the number of 'participating nucleons' $N_{\rm part}$ which are contained in the nuclear overlap volume at a given impact parameter and therefore participate in the collision. 'Centrality' measures the fraction of the total interaction cross-section in bins of a variable which is directly related to the impact parameter; e.g. the '10\% most central events' correspond to the 10\% of events with the highest multiplicity.  Centrality is experimentally measured whereas $N_{\rm part}$ is derived indirectly and in a slightly model dependent 
way.},
$N_{\rm part}$, rises stronger with energy than in pp, approximately with $s^{0.15}$ (Figure~\ref{figpb1} left). Therefore  a central collision between Pb nuclei ($N_{\rm part} \approx 400$) at LHC produces not 200, but 400 times the particle multiplicity of a typical pp collision ($N_{\rm part} = 2$).
Even more surprising is the fact that the centrality dependence of ${\rm d}N_{\rm ch}/{\rm d}\eta$~\cite{Collaboration:2010cz} 
is practically identical to the one of Au-Au at RHIC (at least for $N_{\rm part} >50$), despite the fact that impact parameter dependent nuclear modifications would be expected to be much stronger at LHC. Indeed, models with either strong nuclear modifications or different saturation-type calculations ('Colour Glass Condensate' based models) do describe the impact parameter dependence best.

\begin{figure}[!t]
\centerline{\includegraphics[width=1.0\textwidth]{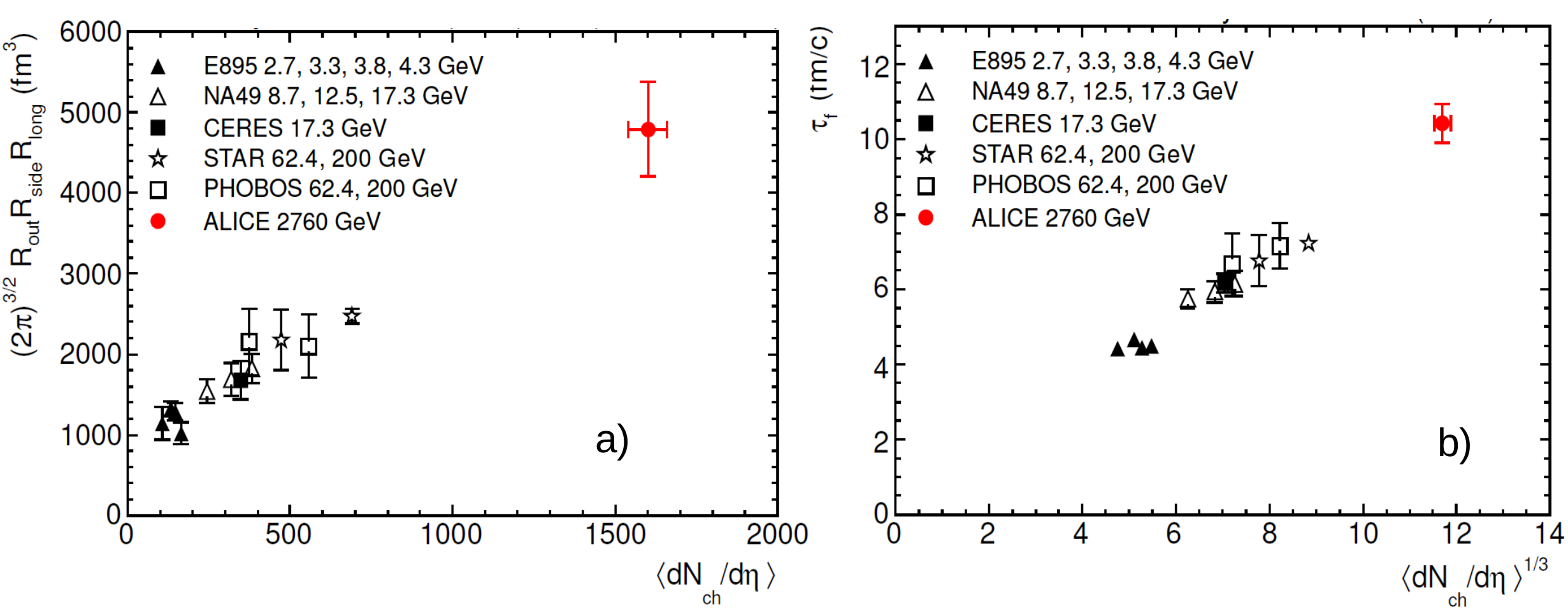}}

\caption{ Left: Local freeze-out volume as measured by identical pion interferometry at LHC compared to central gold and lead collisions at lower energies. Right: The decoupling time $\tau_{f}$ compared to results from lower energies. 
Figures taken from ref~\cite{Aamodt:2011mr}.
}
\label{figpb2}
\end{figure}

The freeze-out volume (the size of the matter at the time when strong interactions cease) and the total lifetime of the created system (the time between collision and freeze-out) was measured with identical particle interferometry (also called Hanbury-Brown Twiss or HBT correlations)~\cite{Aamodt:2011mr}. Compared to top RHIC energy, the local 'comoving' freeze-out volume (Figure~\ref{figpb2}  left) increases by a factor two (to about 5000 $fm^3$) and the system lifetime (Figure~\ref{figpb2}  right) increases by about 30\% (to 10 fm/c), pretty much in line with the predictions of the HISM.  

The most critical test of the heavy ion standard model however comes from the measurement of the elliptic flow\footnote{Flow 
refers to a correlation between space and momentum variables, i.e. particles close in space show similar velocity in both magnitude and direction. Such a collective behaviour is in contrast to random thermal motion where space and momentum variables are in general not correlated. Different flow patterns are observed in heavy ion collisions and quantified in terms of a Fourier decomposition. The second harmonic is called 'elliptic' flow and the zero order (i.e. isotropic) harmonic is called 'radial' flow. Other harmonic components (e.g. 1st, 3rd,..), can be measured as well.}
at LHC, the pillar which supports the 'fluid' interpretation of the QGP. Collective observables like particle flow 
are usually described in the framework of hydrodynamic models, where initial conditions (e.g. geometry and pressure gradients) and fluid properties (e.g. viscosity and equation-of-state) fully determine the observed pattern of collective motions. Assuming  only small or no changes in the fluid properties between RHIC and LHC, hydrodynamic models predict firmly that the elliptic flow coefficient $v_2$, measured as a function of $p_T$, should change very little. As shown in Figure~\ref{figpb1} (right), this prediction was confirmed very quickly and with good precision~\cite{Aamodt:2010pa}. The $p_T$ integrated flow values however do increase compared to RHIC by some 30\%, because the average $p_T$ is significantly higher at LHC. While the average $p_T$ increases also in pp with energy, because hard  and semi-hard processes become more important, hydrodynamics predicts in addition an increase in the radial flow velocity leading to a characteristic $p_T$ and mass dependence of the spectra. It will be very interesting to see, once identified particle spectra will be available, if this prediction is also borne out (as everyone assumes it is), and if the radial flow regime extends to even higher momentum than at RHIC (here predictions are less firm).

With the heavy ion standard model having passed its first tests (HBT, elliptic flow $v_2$) with flying colours, the programme of precision measurements is now starting at the LHC. 
For example, a major advance in quantifying the shear viscosity, which at RHIC was measured to be within a factor 2-4 of the quantum limit of a perfect fluid as conjectured by AdS/CFT, will require a better estimate of remaining non-flow contributions, as well as a better constraint on the initial conditions, i.e. the geometry of the collision zone (and its fluctuations) which drive the various flow components. Progress is already being made on both fronts, in particular by looking in more detail at the centrality dependence and at higher Fourier components ($v_3, v_4,..$).

The energy advantage of the LHC is most evident in the area of parton energy loss (or jet quenching, the 'opaque' aspect of the sQGP), where the kinematic reach vastly exceeds the one available at RHIC. With high $p_T$ jets easily visible above the soft background, jet quenching is qualitatively evident already by visual inspection of charged jets in the TPC where a striking jet energy imbalance develops for central collisions. A quantitative analysis will, however, take more time, in order to see how much energy is actually lost (i.e. measure the transport coefficient), how and where it is lost (multiple soft versus few hard scatterings), and if there is a response of the medium to this local energy deposit (shock waves, Mach cones). Eventually, an answer to all these questions will require comparing in detail how a fast parton fragments into hadrons when travelling through the medium rather than through the vacuum (as in pp collisions), down to low ($<$ 5 GeV) or even very low ($<$ 2 GeV) transverse momentum of the fragments.

 In the meantime, ALICE has measured the nuclear modification of charged particle momentum distributions out to 20 GeV~\cite{Aamodt:2010jd}, where the spectra are dominated by leading jet fragments. The change of the $p_T$ spectrum is most pronounced at around 6 GeV, where the suppression of high momentum particles is modestly stronger than at RHIC, but then rises again smoothly towards higher momentum. This latter feature is not evident in published RHIC data, and while such a rise was qualitatively predicted by some models for LHC~\cite{Abreu:2007kv}, it looks stronger at first sight. However, initial state effects (shadowing/saturation), which presumably are very strong at LHC and which should depend on both impact parameter and momentum transfer, can complicate a straight forward interpretation of the data and the comparison between different beam energies. It will be interesting to see how this result will fit into the overall picture of jet quenching.

All of the heavy ions results summarized above have emerged within weeks of the first Pb-Pb collisions at LHC. They paint a picture which is broadly consistent with the expectations based on lower energy results and confirm that ultra-relativistic heavy ion physics has reached a remarkable level of stability and predictability. Since then, the analysis has progressed significantly and a wealth of new and more detailed results can be expected for the summer conferences of this year (2011). However, it should be kept in mind these are still very early days for heavy ions at the LHC and surprises and unexpected results may yet turn out to require new ingredients 'Beyond the Heavy Ion Standard Model'.

\section{Summary}
After two decades of design, R\&D, construction, installation, commissioning and simulations, the ALICE experiment has had a remarkable smooth and efficient start-up since LHC came into operation at the end of 2009. Most detector systems are fast approaching design performance, and physics analysis is well underway. With the first heavy ion collisions, a new era has started for ultra-relativistic heavy ion physics and one can look forward to at least a decade of exciting (and hopefully revealing) new results concerning matter and the strong interaction under conditions of extreme temperature and density.


\begin{footnotesize}

\end{footnotesize}


\end{document}